\documentclass[aps,twocolumn,superscriptaddress,nofootinbib,showpacs,amsmath,amssymb,superscriptaddress]{revtex4}
\usepackage{graphicx}
\usepackage{colordvi}
\usepackage[dvips]{color}
\usepackage{amsmath}
\usepackage{bm}
\def\k{{\bf k}}

\def\q{{\bf q}}
\begin{document}
\title {Nodal Quasiparticle Lifetimes in Cuprate Superconductors}

\author{T.~Dahm}
\email{thomas.dahm@uni-tuebingen.de}
\affiliation{Institut f\"{u}r
Theoretische Physik, Universit\"at T\"ubingen, T\"ubingen,
Germany}

\author{P.J.~Hirschfeld}
\email{pjh@phys.ufl.edu}
\affiliation{Physics Department,
University of Florida, Gainesville, FL 32611 USA}

\author{D.J.~Scalapino}
\email{djs@vulcan2.physics.ucsb.edu}
\affiliation{Department of
Physics, University of California, Santa Barbara, CA 93106-9530
USA}

\author{L.~Zhu}
\email{zly@phys.ufl.edu} \affiliation{Physics Department,
University of Florida, Gainesville, FL 32611 USA}

\date{\today}
\begin{abstract}

A new generation of angular-resolved photoemission spectroscopy
(ARPES) measurements on the cuprate superconductors offer the
promise of enhanced momentum and energy resolution.  In
particular, the energy and temperature dependence of the on-shell
nodal $(k_x=k_y)$ quasiparticle scattering rate can be studied. In
the superconducting state, low temperature transport measurements
suggest that one can describe nodal quasiparticles within the
framework of a BCS $d$-wave model by including forward elastic
scattering and spin-fluctuation inelastic scattering.  Here, using
this model, we calculate the temperature and frequency dependence
of the on-shell nodal quasiparticle scattering rate in the
superconducting state which determines the momentum width of the
ARPES momentum distribution curves. For a zero-energy
quasiparticle at the nodal momentum $k_N$, both the elastic and
inelastic scattering rate show a sudden decrease as the
temperature drops below $T_c$, reflecting the onset of the gap
amplitude. At low temperatures the scattering rate decreases as
$T^3$ and approaches a zero temperature value determined by the
elastic impurity scattering. For $T>T_c$, we find a quasilinear
dependence on $T$.  At low reduced temperatures, the elastic
scattering rate for the nodal  quasiparticles exhibits a
quasilinear increase at low energy $\omega$ which arises from elastic
scattering processes.  The inelastic spin-fluctuation scattering
leads to a low energy $\omega^3$ dependence which, for
$\omega\gtrsim3\Delta_0$, crosses over to a quasilinear behavior.

\end{abstract}


\maketitle

\section{Introduction}

A $d$-wave BCS framework has proved useful in describing the thermal
\cite{Zha01, Chi99} and microwave \cite{Hos99} conductivities of
the superconducting cuprates \cite{HPS94, WS00, DL00}.  This
suggests that a quasiparticle description of the nodal excitations
is adequate, at least at low energies.  Here, the frequencies are
small compared with the gap magnitude and the temperature
dependence of the nodal quasiparticle lifetime plays the dominant
role. Recent angle-resolved photoemission spectroscopy (ARPES)
experiments on Bi$_2$Sr$_2$Ca$_1$Cu$_2$O$_8$ (BSCCO) are providing
high resolution data on the momentum and energy dependence of the
nodal quasiparticle lifetimes as well as their temperature
dependence in this low energy range \cite{Despc,Val00,Kam00,Kor04}. 
Thus, the question arises whether transport and ARPES data can be 
understood within a single framework. Establishing this 
agreement is important to confirming
the nature of the superconducting phase as a BCS-like
$d$-wave state with nodal quasiparticle excitations.

The width $\Delta k(\omega, T)$ of the momentum
distribution curve (MDC) measured by ARPES experiments is proportional to
the inverse of the on-shell quasiparticle lifetime. Along a nodal
cut in momentum, with $k_x=k_y$, the magnitude $k=\sqrt{k^2_x +
k^2_y}$ is set by $\omega=\varepsilon_k$, i.e. $k$ becomes
effectively a function of $\omega$; we refer to this as
``on-shell".  Here, $\varepsilon_k$ is the quasiparticle energy.
Thus, the lifetime of such a nodal quasiparticle depends upon
$\omega$ and the temperature $T$. As $k$ decreases from the nodal
Fermi momentum $k_N$, $\omega$ can be swept over an energy range
which is significantly larger than $\Delta_0$.  Therefore, there
is a need to extend the inelastic scattering lifetime calculations
to cover a wider range of energies.  In addition, impurity
scattering, particularly due to out-of-plane forward scattering,
is believed to be important so that it is also of interest to
determine its temperature and energy dependence \cite{AV00,ZHS04}.

In the following, we will find it convenient to discuss the elastic 
and inelastic quasiparticle scattering in terms of a scattering rate
\begin{equation}
\Gamma (\omega, \k, T) = - \Sigma^{\prime\prime} (\omega, \k, T)\, ,
\label{one}
\end{equation}
where $\Sigma^{\prime\prime}$ is the imaginary part of the
quasiparticle self-energy.
The inverse of the quasiparticle lifetime  $1/\tau(\omega, \k, T)
= 2 \Gamma (\omega, \k, T)$  and the width of the MDC about
a given $\k$ determined from $\varepsilon_k=\omega$ is
\begin{equation}
\Delta k (\omega, T) = \frac{2\Gamma (\omega, \k|_{\omega=\varepsilon_\k}, T)}
{v(\k)}\, ,
\label{two}
\end{equation}
with $v(\k)=\frac{\partial \varepsilon_\k}{\partial \k}$ the bare band
velocity. Here, we report on results obtained within a $d$-wave BCS framework
for the energy and temperature dependence of the nodal
quasiparticle elastic and inelastic scattering rates and discuss the
$\omega$ and $T$ dependence of $\Gamma (\omega, \k|_{\omega=\varepsilon_k}, T)$.

We begin in Section II with a discussion of the temperature and
energy dependence of the elastic impurity scattering contribution.
In BSCCO, there are both in-plane and out-of-plane scattering
centers \cite{Eis04}. The in-plane impurities are believed to give
rise to strong scattering and are often treated in the unitary
limit \cite{Hud03}.  On the other hand, the out-of-plane impurity
scattering is weak and tends to be forward. It will be treated
within a self-consistent Born approximation. We will see that the
temperature dependence of the elastic scattering rate arises from
the opening of the gap as $T$ decreases below $T_c$, while its
energy dependence reflects the decreased phase space as $\omega$
becomes smaller than the gap magnitude.  The low energy inelastic
scattering discussed in Section III is dominated by short range
Coulomb scattering.  At low energies, the linear $\omega$
dependence of the density of states gives rise to an inelastic
scattering rate which varies as $T^3$ for $T$ larger than $\omega$
or $\omega^3$ for $\omega$ larger than $T$.  At higher energies,
one can probe the $\omega$-dependence of the effective interaction.
Here, we confine ourselves
to the Hubbard model and explore the spin-fluctuation
contributions to the higher energy behavior of the inelastic
scattering. In Section IV, we discuss the
combined effects of the elastic and inelastic scattering
processes, and Section V contains our conclusions.

\section{Elastic Scattering}

Elastic scattering in BSCCO can arise from both impurities and
disorder in the CuO planes as well as from regions outside these
planes. The  scattering rate due to   unitary scatterers (possibly
Cu vacancies) with concentration roughly $n_u\sim 0.2\%$ observed
as zero-bias resonances in scanning tunneling microscopy (STM) 
experiments \cite{Hud03} is well
understood theoretically. It will be treated as usual in the
self-consistent $T$-matrix approximation,
\begin{equation}
\underline{\Sigma}_{el,u} = - \frac{n_u}{\sum_{\bf k} G({\bf
k},\omega)}\, \tau_0, \label{three}
\end{equation}
where $\tau_i$ are the $2 \times 2$ Pauli matrices in
Nambu notation.
At low frequencies and temperatures, the unitary scatterers give
rise to an impurity resonance of height
$\Sigma_{el,u}''(\omega=0)\equiv \gamma_u\simeq \sqrt{\Gamma_u
\Delta_0} \sim 10^{-2}t$, and roughly the same width, 
with $t$ the nearest-neighbor hopping matrix element. In the
normal state we find a scattering rate of $\Gamma_u \sim n_u
E_F\simeq 10^{-3}t$ due to these defects, leading to $\gamma_u
\simeq$ 1 to 2 meV.  If, as we believe, the normal state elastic
scattering rate at the node is a significant fraction of the gap,
the contribution of the unitary scatterers will be difficult to
resolve in ARPES experiments on BSCCO.

The elastic scattering from out-of-plane impurities, such as
interstitial O ions, can be modeled by a momentum-dependent
potential,
\begin{equation}
V(\k, \k^\prime)=V_0 f(\k, \k^\prime) \label{four}
\end{equation}
which has a forward scattering form factor $f(\k, \k^\prime)$ which
cuts off the scattering when $|\k-\k^\prime|$ exceeds a momentum
$\kappa$, as has been discussed in Ref.~\onlinecite{ZHS04}. 
We will measure $\kappa$ in units of $k_F$ so that
$\kappa^{-1}$ characterizes the range of the impurity potential in
units proportional to the Cu-Cu spacing $a$. Since the scattering
from an out-of-plane impurity is relatively weak, it can be
treated within the self-consistent Born approximation \cite{Sch05}.  For a
nodal quasiparticle with $k_x=k_y$, the gap vanishes and the
scattering rate due to elastic scattering is determined from the
imaginary part of the $\tau_0$ Nambu self-energy in the
self-consistent Born approximation,
\begin{equation}
\Sigma_{el,0} (\k, \omega) = n_i \sum_{k^\prime} |V(\k,
\k^\prime)|^2 \ \frac{\tilde\omega}{\tilde\omega^2 -
\varepsilon^2_{k^\prime} - \tilde\Delta^2_{k^\prime}},
\label{five}
\end{equation}
where $n_i$ is the planar density of the out-of-plane impurities.
In addition, because the potential $V(\k, \k')$ is anisotropic, it
is necessary to renormalize the order parameter as well:
\begin{equation}\Sigma_{el,1} (\k,\omega)= n_i \sum_{k^\prime} |V(\k,
\k^\prime)|^2 \ \frac{\tilde\Delta_{\k'}}{\tilde\omega^2 -
\varepsilon^2_{k^\prime} - \tilde\Delta^2_{k^\prime}},
\label{six}
\end{equation}
where $\tilde\omega = \omega-\Sigma_{el,0}$ and $\tilde \Delta_\k
= \Delta_\k +\Sigma_{el,1}$.  The renormalization of $\varepsilon_\k$
by disorder is also nonzero in general but was shown to be
negligible in Ref.~\onlinecite{ZHS04}, and will therefore be
neglected here.
  For our numerical calculations we will
use a simple parameterization of the band structure with
\begin{equation}
\varepsilon_k=-2t\, (\cos k_x+\cos k_y) - 4t^\prime \cos k_x \cos k_y - \mu
\label{seven}
\end{equation}
and
\begin{equation}
\Delta_k = \frac{\Delta_0}{2}\ (\cos k_x-\cos k_y)\, .
\label{eight}
\end{equation}
Here, $t^\prime = -0.35$ and $\mu=-1.1$, with energy measured in units of
nearest neighbor hopping $t$. For the temperature dependence of the
gap we use a common interpolation form
\begin{equation}
\Delta_0(T) = \Delta_0 \tanh
\left(\alpha\sqrt{\frac{T_c}{T}-1}\right) \label{gapinterp}
\end{equation}
with $\alpha=3$, $2\Delta_0/k\, T_c=6$, and $\Delta_0=0.2$.
These parameters are chosen
to give the observed magnitude of the $T=0$ gap and a somewhat more rapid
rise at $T_c$, consistent with experiment.

\begin{figure}[t]
\leavevmode
\includegraphics[width=.6 \columnwidth,clip,angle=0]{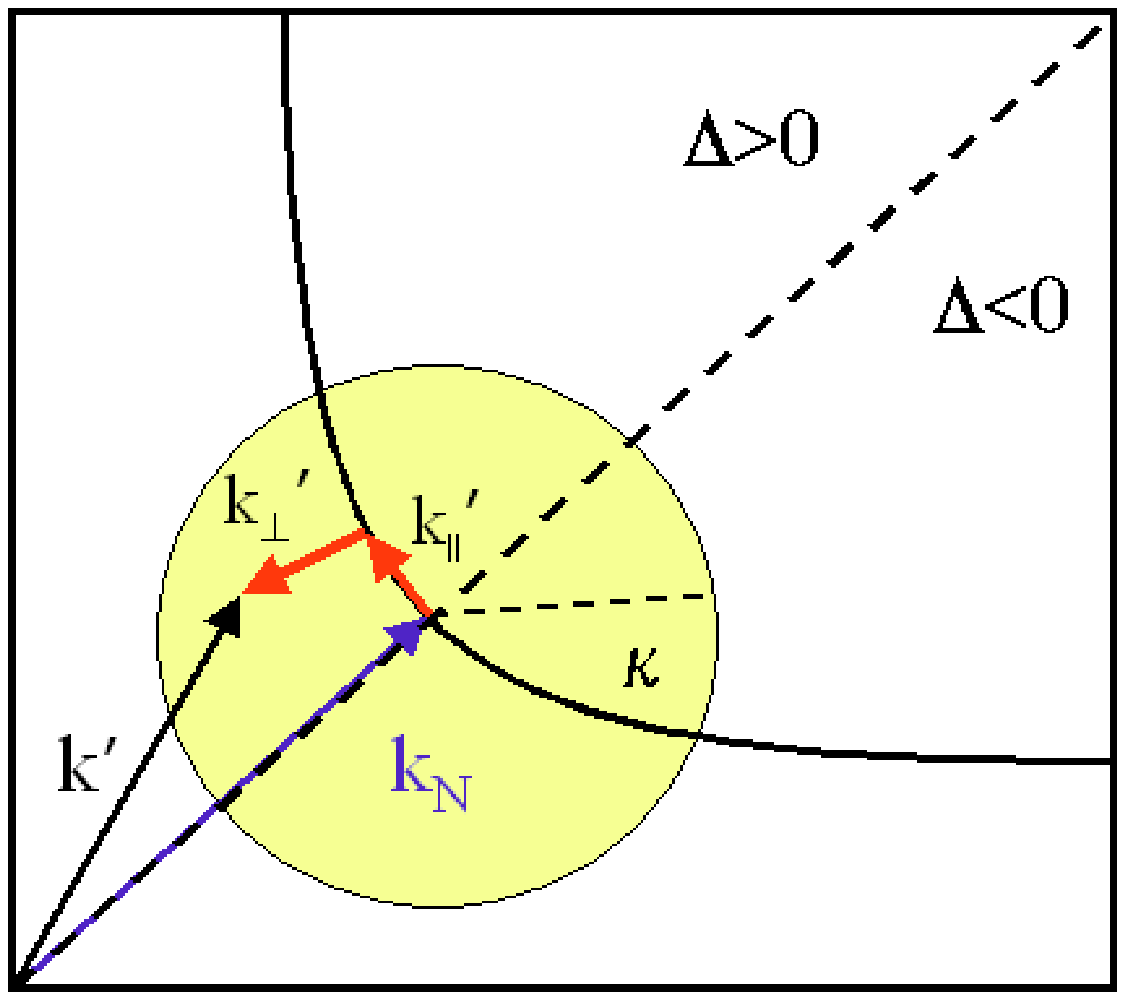}
\caption{(Color online) The Fermi surface corresponding to the band 
$\varepsilon_k$
given by Eq.~\eqref{seven} with $t^\prime/t=-0.3$ and $\mu=-1.1$.
The difference of the scattered wave vector $\k'$ from the nodal
wave vector $\k_N$ is shown as its parallel displacement along the
Fermi surface $\k_\parallel$ and the perpendicular displacement
$\k_\perp$. } \label{fig:FS}
\end{figure}

The scattering rate $\Gamma_{el} (\k_N, \omega)$ is now determined by
the imaginary part of the solution of Eq.~\eqref{five} at the
nodal wavevector $\k_N$. Before turning to a numerical evaluation
of Eq.~\eqref{five}-\eqref{six}, we consider a simple analytic
approximation similar in spirit to that discussed in \cite{Kee01}
in the context of $T_c$ suppression by disorder \cite{Tcsupress}. Measuring the
momentum transfer from ${\bf k}_N$ along the $k_\perp$ and
$k_\|$ coordinates shown in Fig.~\ref{fig:FS}, the imaginary part of
Eq.~\eqref{five} can be written as
\begin{equation}
\Gamma_{el}  = - n_i |V_0|^2\, {\rm Im} \int  \frac{d k^\prime_\|
d\, \k^\prime_\perp}{(2\pi)^2}\ | f(\k, \k^\prime)|^2\
\frac{\tilde\omega}{\tilde\omega^2-v^2_1\, k'^2_\perp - \tilde
v^2_2\, k'^2_\|}
\label{ten}
\end{equation}
Here, $v_1=v_F$ at the nodal point and $\tilde v_2$ is the renormalized gap
velocity. We have assumed that the
$\kappa$ cut-off prevents scattering to other nodes.

 Since $v_1\gg v_2$, the $\kappa$ cut-off primarily
affects the $k_\|$ integral and one can integrate freely over the
important range $k_\perp$ giving
\begin{equation}
\Gamma_{el} = \frac{1}{2} n_i\, N_0 |V_0|^2 {\rm Re} \int^{\kappa}_{-\kappa} d
k'_\| \ \frac{\tilde\omega}{\sqrt{\tilde\omega^2 - \tilde v^2_2 \,
k'^2_\|}}
\label{eleven}
\end{equation}
with $N_0=(2 \pi\, v_F)^{-1}$ the band density of states in units
in which $k_F=1$. Setting $\tilde\omega=\omega+i\, \Gamma_{el}$,
i.e. neglecting the real part of the self-energy, this becomes
\begin{equation}
\Gamma_{el} =\frac{\Gamma_0}{\tilde v_2\kappa}\ {\rm
Re}\left[(\omega+i\, \Gamma_{el}) \sin^{-1} \left(\frac{\tilde v_2
\kappa}{\omega+i\, \Gamma_{el}}\right)\right]\, ,
\label{twelve}
\end{equation}
where $\Gamma_0=n_i\, N_0\, |V_0|^2\kappa$ is
the normal state elastic scattering rate and $\tilde v_2\kappa$ is
the maximum  renormalized order parameter  probed by the
scattering.  From this result we recover immediately the
$\omega\rightarrow 0$, $k_\|\rightarrow 0$ limit that
$\Gamma_{el}\rightarrow \Gamma_0$ from Ref.~\onlinecite{ZHS04}.

\begin{figure}[t]
\leavevmode
\includegraphics[width=.7 \columnwidth,clip,angle=-90]{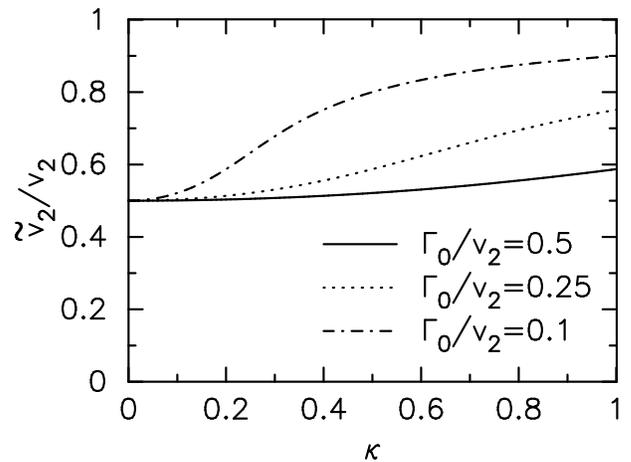}
\caption{Gap slope renormalization $\tilde v_2 /v_2$ vs.~inverse
scattering range $\kappa$ from the solution of Eqs.~\eqref{twelve} and
\eqref{fourteen}. } \label{fig:vtilde}
\end{figure}

The gap velocity at the node in the pure system, $v_2\simeq
2\Delta_0/k_N$, is significantly renormalized in the presence of
disorder to $\tilde v_2$. From Eq.~\eqref{six} we find, for
$k_\|\ll\kappa$,
\begin{eqnarray}
(\tilde v_2 -v_2) k_\| & = & \frac{\Gamma_0}{2\kappa}\, {\rm Im}
\int^{\kappa}_{-\kappa} {d k^\prime_\|} \ \ \frac{\tilde v_2
(k_\|'+k_\|)}{\sqrt{\tilde\omega^2 - \tilde v^2_2\,
(k'_\|+k_\|)^2} }
\label{thirteen}
\end{eqnarray}
leading to
\begin{eqnarray}
\tilde v_2 -v_2 & \simeq & -{\Gamma_0} \frac{\tilde v_2}{
\sqrt{\Gamma_{el}^2+ \tilde v^2_2\kappa^2}}.
\label{fourteen}
\end{eqnarray}
This result, which should be solved self-consistently together with
Eq.~\eqref{twelve}, obtains only in the limit where the
scattering is sufficiently forward such that the momentum
integration takes place only near the node; it interpolates
correctly to the isotropic limit $\kappa\rightarrow \infty$,
however, where the gap velocity renormalization vanishes, $\tilde
v_2 = v_2$.  In the forward scattering limit $\kappa \rightarrow
0$, it is easy to check that $\tilde v_2\rightarrow v_2/2$, and
the dependence for small $\kappa$ is otherwise weak, as shown in
Fig.~\ref{fig:vtilde}.

\begin{figure}[t]
\begin{center}
\leavevmode
\includegraphics[width=0.7 \columnwidth,clip,angle=-90]{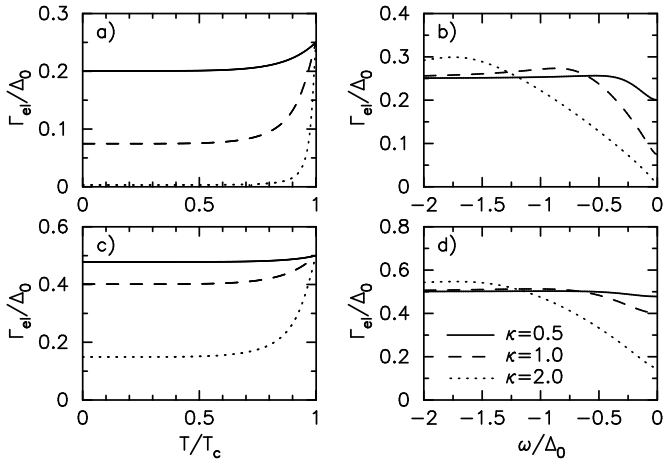}
\caption{Evaluation of the elastic nodal scattering rate
$\Gamma_{\rm el}$ from Eqs.~\eqref{twelve} and \eqref{fourteen}.  a)
temperature dependence of
 $\Gamma_{\rm el}(\omega=0,T)$  for three
values of the impurity potential range parameter
$\kappa=0.5$ (solid), 1.0 (dashed), and 2.0 (dotted) 
with $\Gamma_0=0.25\Delta_0$; b) energy
dependence of $\Gamma_{\rm el}(\omega,T=0)$ for the same parameters; c),d)
same as a),b) but for $\Gamma_0=0.5\Delta_0$.}
\label{fig:forwardapprox}
\end{center}
\end{figure}

We will consider two different ways of characterizing the scattering
rates of quasiparticles with $k_x=k_y$. First, suppose $\k$ is
fixed exactly at the nodal Fermi surface $\k_{N}$.  In this case
the quasiparticle energy $\omega=0$ and one can study
$\Gamma_{el}(\k_{N}, \omega=0)$ as a function of temperature.  The
contribution of the out-of-plane elastic impurity scattering to
the temperature dependence of the MDC line width $\Delta k(T)$ for
a quasiparticle with $\k=\k_{N}$ is given by $2\Gamma_{el}(\k_{N},
\omega=0)/v_F$. When $\omega=0$, Eq.~\eqref{twelve} can be written
as
\begin{equation}
1=\frac{\Gamma_0}{\tilde v_2\kappa}\ \ln
\left(\frac{1+\sqrt{1+(\Gamma_{el}/\tilde
v_2\kappa)^2}}{(\Gamma_{el}/\tilde v_2\kappa)}\right)\, .
\label{fifteen}
\end{equation}
At low temperatures, if $\tilde v_2 \kappa/\Gamma_0\gg 1$, we find  the
familiar Born result
\begin{equation}
\Gamma_{\rm el}(T) \simeq \tilde v_2\kappa\, e^{-\left(\frac{\tilde
v_2\kappa}{\Gamma_0}\right)}
\label{sixteen}
\end{equation}
with the gap maximum $\Delta_0$ in the isotropic case replaced by
$\tilde v_2\kappa$. When $\tilde v_2 \kappa/\Gamma_0 \ll 1$,
\begin{equation}
\Gamma_{\rm el} (T) \simeq \Gamma_0\left(1-\frac{1}{6}
\left(\frac{\tilde v_2\kappa}{\Gamma_0}\right)^2\right) \, .
\label{seventeen}
\end{equation}
This limit, of course, always applies when $T$ goes to $T_c$, but it
can also apply for all values of $T/T_c$ if the system is
sufficiently dirty. Note that since $\Gamma_0$ varies as $|V_0|^2
\kappa$, if $V_0$ is kept fixed, the normal state scattering rate
decreases as $\kappa$ becomes smaller. This simply reflects the
reduction of phase space as the scattering is restricted to be in
a more narrow forward cone.  At the same time, the drop in the
scattering rate as the gap $\Delta_0(T)$ opens for $T<T_c$, is
smaller for reduced values of $\kappa$.  This reflects the fact
that the maximum effective gap reached by a scattered nodal
quasiparticle is reduced by $\kappa$ so that the opening of the
gap is less effective in suppressing the scattering rate. Results
obtained from solving Eqs.~\eqref{twelve} and \eqref{fourteen} for
$\Gamma_0/\Delta_0=0.5$ and 0.25 are shown in
Fig.~\ref{fig:forwardapprox} a) and c) as a function of $\kappa$.
The temperature dependence of the elastic scattering arises from
the temperature dependence of the $d$-wave gap, and therefore
varies as $T^3$ at low $T$. Note that the interpolation form,
Eq.~\eqref{gapinterp}, actually gives a qualitatively incorrect
result at low temperatures for a $d$-wave superconductor.
Eq.~\eqref{gapinterp} implies an activated thermal depletion of
the condensate, whereas the correct solution to the $d$-wave gap
equation has a leading term varying as $T^3$
\cite{dwavegapTdependence}, giving $\Delta_0(T) \simeq \Delta_0
(1-bT^3/\Delta^3_0)$, where $b$ is a constant of order unity. In
the clean limit, it follows from Eq.~\eqref{sixteen} that
\begin{eqnarray}
\Gamma_{\rm el} (T)&\simeq&  \Gamma_{\rm el} (T=0)\,\left(
1+\frac{b(\tilde v_2\kappa -\Gamma_0)}{\Delta_0^3 \Gamma_0}\,T^3
\right)\nonumber\\ &\simeq& \Gamma_{\rm el} (T=0)\,
\left(1+2\frac{b\kappa T^3}{\Gamma_0\Delta^2_0}\right)
\label{ninea}
\end{eqnarray}
while in the dirty limit, Eq.~\eqref{seventeen} implies that
\begin{eqnarray}
\Gamma_{\rm el} (T)&\simeq& \Gamma_0\left(1-\frac{(\tilde
v_2\kappa)^2}{ 6 \Gamma_0^2} \right)+
\frac{b(\tilde v_2 \kappa)^2}{3 \Delta_0^3\Gamma_0} T^3\nonumber\\
&\simeq& \Gamma_0 \left(1 -\frac{\Delta_0^2\kappa^2}{ 6 \Gamma_0^2}
\right) + \frac{b\kappa^2}{3 \Delta_0\Gamma^2_0}T^3\, .
\label{nineb}
\end{eqnarray}
In Eqs. \eqref{ninea} and \eqref{nineb} we have assumed $v_2
\simeq 2 \Delta_0$ and used the asymptotic results for $\tilde
v_2/v_2$ derived above to obtain simple expressions for the clean
and dirty limits, respectively.

It is also interesting to study the $\omega$ energy dependence of
$\Gamma_{el}$ at a fixed temperature. Within the current
approximate framework represented by Eqs.~\eqref{twelve} and
\eqref{fourteen}, we will study $\Gamma_{\rm el}(\omega,\k_N,T)$.
For $T$ small compared with $T_c$, one obtains the results shown
in Fig.~\ref{fig:forwardapprox} b) and d). The initial increase in
the scattering rate as the binding energy $|\omega|$ increases
reflects the opening of the nodal phase space with increased
quasiparticle energy. When $|\omega|$ is large compared to $\tilde
v_2\kappa$, the scattering saturates at its normal state value. By
expanding Eq.~\eqref{twelve}, we find that
\begin{equation}
\Gamma_{\rm el} (\omega)\simeq \Gamma_0\left(1+\frac{1}{6}\
\left(\frac{\tilde v_2\kappa}{\omega}\right)^2\right)
\label{eighteen}
\end{equation}
so this saturation actually occurs from values of $\Gamma_{el}$
somewhat above $\Gamma_0$, as seen in
Fig.~\ref{fig:forwardapprox}.

For forward scattering, where $\tilde v_2\kappa/\Gamma_0$ is
small, one can obtain a useful estimate of the quasilinear slope.
In this case, $\Gamma_{\rm el} (\omega)$ rises from its value of
$\Gamma_{\rm el}(\omega=0)$ at $\omega=0$ to $\Gamma_0$ over a frequency range of order $\tilde
v_2\kappa$ with a quasi-linear slope $(\Gamma_0-\Gamma_{\rm el} (\omega=0))/(\tilde v_2\kappa)$. In
the dirty limit, Eq.~\eqref{seventeen}, for $0<|\omega| < \tilde v_2\kappa$ this gives
\begin{equation}
\Gamma_{\rm el}(\omega) \simeq \Gamma_{\rm el} (\omega=0) + \frac{1}{6}\ \left(\frac{\tilde
v_2\kappa}{\Gamma_0}\right)\, \omega \, .
\label{nineteen}
\end{equation}
Finally, if the weak scattering is sufficiently isotropic $(\kappa
\gtrsim 1)$, we recover the standard result for pointlike weak
scatterers
\begin{equation}
\Gamma_{\rm el} (\omega) \simeq \Gamma_{\rm el} (\omega=0) +
\frac{\pi}{2}\ \frac{n_iN_0 |V_0|^2\omega}{\Delta_0}\,.
\label{nineteena}
\end{equation}
Which of the results, Eq.~\eqref{nineteen} or \eqref{nineteena} is valid depends upon the
character of the disorder.

\begin{figure}[t]
\begin{center}
\leavevmode
\includegraphics[clip=true,width=1.0\columnwidth]{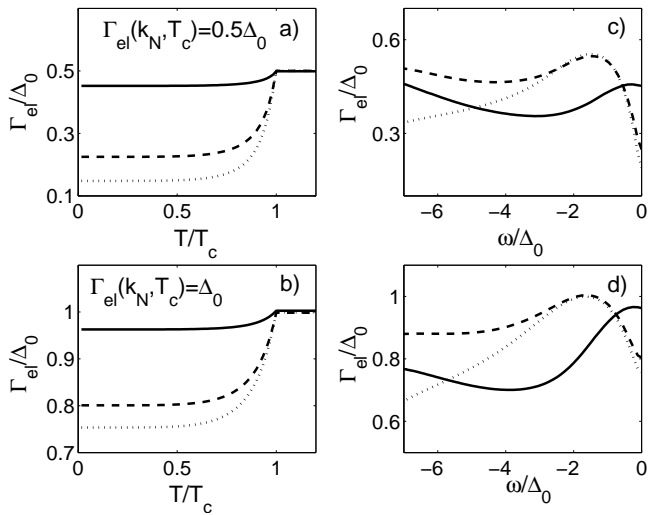}
\caption{Elastic scattering rate $\Gamma_{\rm el}$ for the
Yukawa-type potential with range $\kappa^{-1}$ Eq.~(\ref{twenty}).
a): $\Gamma_{\rm el}/\Delta_0$ vs.~$T/T_c$ for a normal state
elastic scattering rate $\Gamma_{\rm el} (k_N, T_c)=0.5 \Delta_0$
and $\kappa=0.1$,  1, and $\infty$ (isotropic). c): $\Gamma_{\rm
el}/\Delta_0$ vs.~$\omega/\Delta_0$ for the same $\kappa$ and
$\Gamma_{\rm el} (k_N, T_c)$ values. b) and d): same quantities
plotted as above, but with $\Gamma_{\rm el} (k_N, T_c)=\Delta_0$.
Results for different values of the scattering parameter
$\kappa=0.1$ (solid), 1 (dashed), and the isotropic case
$\kappa=\infty$ (dotted) are shown.} \label{fig:Yukawa}
\end{center}
\end{figure}

To check these estimates, and evaluate the
scattering rate for various scattering potential ranges
and impurity concentrations, as well as to treat momenta away from
the Fermi surface more accurately, we introduce a more
realistic model for the elastic scattering. We consider a screened exponential
fall-off in 2D such that,
\begin{eqnarray}
\label{twenty}
  |V(\k,\k')|^2 &=& \frac{|V_0|^2}{\q^2 + \kappa^2},
\end{eqnarray}
where $\q = \k-\k'$ is the momentum transfer.  We then perform a
self-consistent solution of equations \eqref{five} and
\eqref{six}.  Results for $\Gamma_{\rm el}$ analogous to 
Fig.~\ref{fig:forwardapprox}, but taken ``on-shell" at momentum $\k$
such that $\varepsilon_\k=\omega$ are presented in Fig.~\ref{fig:Yukawa}.  
Results have been scaled such that the
(self-consistent) normal state scattering rate is the same for all
curves; as discussed above, this implies that the more forward
scattering cases correspond to larger impurity concentrations.  It
is seen that the qualitative behavior close to the Fermi level is
captured rather well by the approximate model discussed above where the scattering is
restricted to a cone of width $\sim \kappa$. It is also
interesting to note that the quasilinear behavior at the low
$\omega$ values persists until extremely long scattering ranges, of
order 5-10 lattice spacings, for reasonable total normal state
scattering rates.  This suggests on the one hand that elastic
scattering is a likely explanation for the quasilinear behavior
seen at low energies in the MDC width; on the other hand,
it also means that it will be difficult to determine the range of
the scatterers precisely from such a measurement alone.

\section{Inelastic Scattering}

In this section we discuss the scattering rates
due to inelastic electron-electron scattering by a phenomenological
short range
Coulomb interaction based on the two dimensional Hubbard model.
For the tight-binding bandstructure considered here, these
scattering processes are dominated by exchange of
antiferromagnetic spin-fluctuations. We are using a
conventional Berk-Schrieffer-like \cite{Berk} theory as has been
used before to discuss the single-particle
inelastic lifetime \cite{QSB94} and has proved to give a qualitative description of
low-energy microwave and thermal conductivity lifetimes \cite{DSH01,QHS96}
as well as NMR relaxation rates \cite{BS92} in the cuprates.

In second-order perturbation theory for the two dimensional 
Hubbard model, the imaginary part of the
nodal quasiparticle self-energy due to inelastic scattering from
the onsite Coulomb interaction $U$ can be written as
\begin{eqnarray}
\lefteqn{ \Gamma_{\rm inel} (\omega, \k, T) = -\Sigma_{inel}^{\prime\prime}
(\omega, \k, T) = } \nonumber \\
 & & \frac{U^2}{N} \sum_{q} \int^\infty_{-\infty} d\Omega
\, [n(\Omega)+f(\Omega-\omega)]\, \cdot \nonumber \\
& & \chi^{\prime\prime}_0 (q, \Omega) \, N(k-q, \omega-\Omega)
\label{twentyone}
\end{eqnarray}
with
\begin{eqnarray}
N(k, \omega) & = & -\frac{1}{\pi}\ {\rm Im} \left(\frac{\omega+i\delta +
\varepsilon_k}{(\omega+i\delta)^2-E^2_k}\right)
\label{twentytwo}
\end{eqnarray}
and
\begin{eqnarray}
\lefteqn{
\chi_0 ({\bf q}, \Omega) =} \nonumber \\
& & \frac{1}{N} \sum\limits_k \Biggl\{\frac{1}{2}\left[1 +
\frac{\varepsilon_{k+q} \varepsilon_k + \Delta_{k+q}\Delta_k}{E_{k+q}E_k}\right]
\ \frac{f(E_{k+q})-f(E_k)}{\Omega-(E_{k+q}-E_k)+ i0^+} \nonumber\\
& &  + \frac{1}{4}\ \left[1 -
\frac{\varepsilon_{k+q}\varepsilon_k+\Delta_{k+q}\Delta_k}{E_{k+q}E_k}\right]
 \frac{1-f(E_{k+q})-f(E_k)}{\Omega+(E_{k+q}+E_k) + i0^+}\nonumber\\
& & + \frac{1}{4}
\ \left[1-\frac{\varepsilon_{k+q}\varepsilon_k+\Delta_{k+q}\Delta_k}{E_{k+q}E_k}\right]
\ \frac{f(E_{k+q})+f(E_k)-1}{\Omega-(E_{k+q}+E_k)+i0^+}\Biggr\}\, .
\label{twentythree}
\end{eqnarray}
Here, $E_k=\sqrt{\varepsilon_k^2+\Delta_k^2}$. 
As before, we are considering a nodal quasiparticle with
$k_x=k_y$ and the magnitude of $k$ is set by
$\omega=\varepsilon_k$.  At low energies, where $\omega$ and $T$
are small compared to the zero temperature gap amplitude, Eq.~\eqref{twentyone}
determines the form of the
$\omega$ and $T$ dependence of the inelastic quasiparticle
scattering.  Higher order processes lead to a renormalization of
$U$, but the phase space restrictions imposed by the $d$-wave nature
of the superconducting state determine the low $T$ and $\omega$
dependence of $\Gamma_{\rm inel} (\omega, \k|_{\omega=\varepsilon_\k},
T)$ \cite{QSB94,analytic,DSH01}.

\begin{figure}[t]
\begin{center}
\leavevmode
\includegraphics[clip=true,angle=-90,width=1.0\columnwidth]{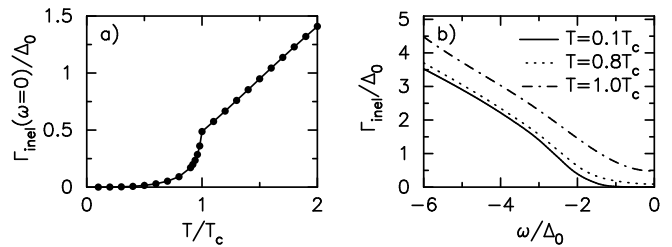}
\caption{Left panel: Temperature dependence of inelastic RPA spin
fluctuation scattering rate $\Gamma_{\rm inel}(0,\k_N,T)/\Delta_0$
vs.~$T/T_c$ for $U=2.2t$.   Right panel: Frequency dependence
$\Gamma_{\rm inel}(\omega,\k|_{\omega=\varepsilon_\k},T)/\Delta_0$
vs.~$\omega/\Delta_0$ for reduced temperatures $T/T_c$=0.1, 0.8, and 1.0.}
\label{fig:inelastic1}
\end{center}
\end{figure}

For $\k=\k_N$ and $\omega=0$, the low temperature inelastic
scattering varies as
\begin{equation}
\Gamma_{\rm inel} (0, T) \simeq  \frac{T^3}{\Delta^2_0}
\label{twentyfour}
\end{equation}
with a prefactor of order one \cite{QSB94,analytic}. For the
set of parameters discussed below the prefactor becomes 2.4.
It has been argued that $\Gamma_{\rm inel}$ should vary
as $T^{5/2}$ at low temperatures \cite{Paaske}.
However, for the temperatures $T>0.05 T_c$ we
have studied, the $T^3$ behavior provides a much better fit
to our numerical results in Fig. \ref{fig:inelastic1}(a).
Similarly, at $T=0$ the $\omega$
dependence is given by
\begin{equation}
\Gamma_{\rm inel} (\omega, 0) \simeq  \frac{\omega^3}{\Delta^2_0}\, .
\label{twentyfive}
\end{equation}
The extra power $T^3$ and $\omega^3$ versus the $T^2$ and
$\omega^2$ dependence of the usual (3D) Fermi liquid inelastic
quasiparticle Coulomb scattering is due to the $\omega/\Delta_0$
dependence of the $d$-wave density of states.  Note that the usual
$T^3$ and $\omega^3$ phonon contributions to the low energy
quasiparticle scattering in a normal metal give contributions
which vary as $T^4$ and $\omega^4$ in a $d$-wave
superconductor.

\begin{figure}[t]
\leavevmode
\includegraphics[width=.7 \columnwidth,clip,angle=-90]{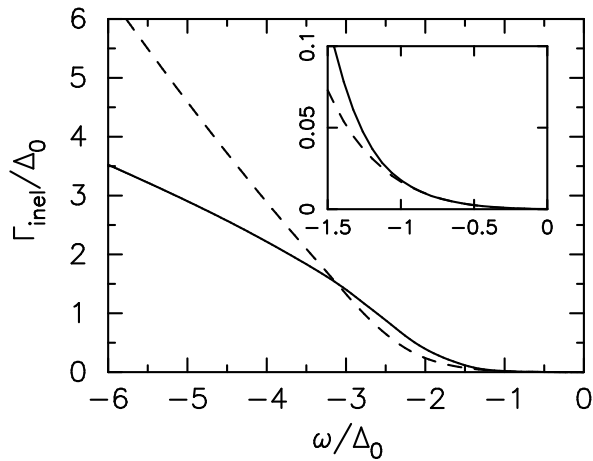}
\caption{Comparison of on-shell spin-fluctuation inelastic scattering rates
for $\Gamma_{inel}(\omega,\k|_{\omega=\varepsilon_\k},T)/\Delta_0$
vs.~$\omega/\Delta_0$ for a reduced temperature $T/T_c=0.1$ .  Dashed line:
2nd order perturbation theory with  $U=6.7t$.  Solid line:
RPA theory with $U=2.2t$.  Inset shows magnified low energy
region. } \label{fig:2ndordervsRPA}
\end{figure}

At higher energies, the problem becomes more complicated.  
New, collective channels may open
such as the so-called $\pi$-resonance or possibly the $B_{1g}$ phonon. 
Here, we replace $\chi_0(q,\omega)$ by the RPA form
\begin{equation}
\chi(q,\omega)=\frac{\chi_0(q, \omega)}{1-U\chi_0(q, \omega)}
\label{twentysix}
\end{equation}
and consider the spin-fluctuation contributions to the scattering
of the nodal quasiparticle. Following the usual spin-fluctuation 
notation, we replace the coupling $U^2$ in Eq.~\eqref{twentyone} 
by $\frac{3}{2} U^2$ which in any case is simply a
phenomenological coupling constant in these calculations.
In principle, the $U$ that enters
the coupling is different from the effective $U$ in the
denominator of $\chi$, Eq.~\eqref{twentysix} because of vertex
corrections. Here, we will for
simplicity ignore this distinction so we have just the basic RPA
form of the spin-fluctuation interaction. A similar approach was
used to discuss the microwave and thermal conductivity lifetimes
\cite{DSH01}. This approach will in principle include resonant
spin excitations like those which have been proposed to explain
the $\pi$ resonance, i.e. the peak seen by neutron scattering at
around 40 meV in various cuprates, but we do not investigate these
effects in detail here. Rather, we will examine what this
approximation gives for the higher energy dependence of
$\Gamma_{\rm inel} (\omega, \k|_{\omega=\varepsilon_\k}, T)$
numerically.

In Fig.~\ref{fig:2ndordervsRPA}, we show the numerical
evaluation of Eq.~\eqref{twentyone},
as well as the RPA result with
$\chi_0$ replaced by $\chi$ as in Eq.~\eqref{twentysix}, using the band
parameters discussed above in Eqs.~\eqref{seven}
and \eqref{eight} appropriate for an optimally doped
cuprate. Here, the numerical integrations in Eq.~\eqref{twentyone}
and \eqref{twentythree}
are done using the technique described
in Ref.~\cite{DT95}.  The value of $U=2.2t$ chosen in the
RPA case comes from early fits of spin-fluctuation theory to NMR
data, and more recently to microwave and thermal conductivity data
on YBCO, but is expected to be reasonable for BSCCO as well
because transport data suggest that the inelastic scattering rates at $T_c$
are very similar.  The value of $U=6.7t$ in the case of the second
order perturbation theory result was chosen so that the
$\omega^3$ behavior of both results matches for
$|\omega|/\Delta_0 < 1$ as shown in the inset.
As noted, when $|\omega| \lesssim \Delta_0$, the
$\omega$-dependence of the inelastic scattering is determined by the phase space,
and one gets an $\omega^3$ dependence from a constant $U$ interaction as well as from the spin
fluctuation (RPA) interaction. However,
as shown in Fig.~\ref{fig:2ndordervsRPA}, 
at higher energies the scattering reflects the frequency
dependence of the effective interaction.
In the RPA case,
the classical quasilinear energy dependence expected in spin
fluctuation theory when the energy exceeds the spin-fluctuation
energy is clearly visible, and the crossover to the low-energy
$\omega^3$ form takes place around $3\Delta_0$, as also
found in Ref.~\cite{QHS96}.

In Figures \ref{fig:inelastic1} a) and b), we study the effect of
temperature on the inelastic scattering.  First, in
Fig.~\ref{fig:inelastic1} a) we consider the nodal scattering rate at
$\omega=0$, which collapses rapidly at $T_c$ due to the
opening of the gap.  In Fig.~\ref{fig:inelastic1} b), the energy
dependence of the scattering rate is plotted for various
temperatures.  It is interesting to note that for these band
parameters, there is a considerable amount of upward curvature at
low $\omega$ in $\Gamma_{\rm el}(\omega,\k|_{\omega=\varepsilon_\k},T)$ even at $T_c$, in
contrast to the simpler nearly nested bands considered in
\cite{QSB94,QHS96}.

\section{Total scattering rate}

\begin{figure}[t]
\begin{center}
\leavevmode
\includegraphics[clip=true,angle=0,width=1.0\columnwidth]{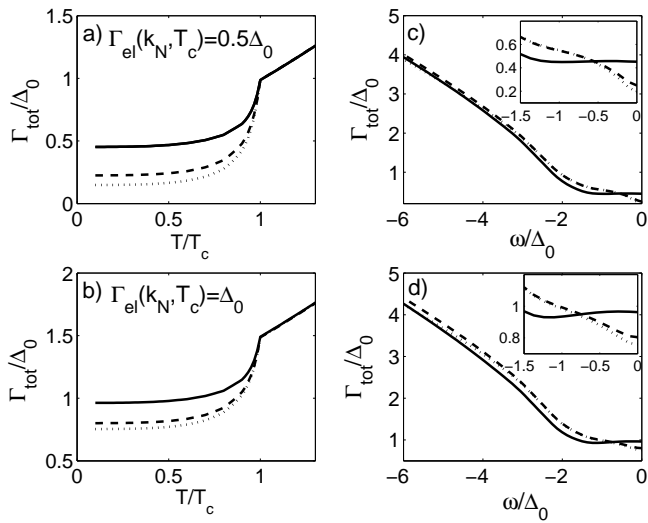}
\caption{a) and b): Temperature dependence of the total scattering
rate $\Gamma_{\rm tot} (0, k_N, T)/\Delta_0$ vs.~$T/T_c$ for
$U=2.2t$ and $\Delta_0/t=0.2$, and normal state elastic scattering
rates of $\Gamma_{\rm el} (0, k_N, T_c)/\Delta_0=0.5$ and 1,
respectively. c) and d): Frequency dependence of the on-shell
scattering rate $\Gamma_{\rm tot}(\omega,
k|_{\omega=\varepsilon_k})/\Delta_0$ vs. $\omega/\Delta_0$ for the
same $U$ and $\Delta_0$ and $T=0.1 T_c$ with $\Gamma_{\rm el}(0,
k_N, T_c)/\Delta_0=0.5$ and 1, respectively.  
Results for different values of
$\kappa=0.1$ (solid), 1 (dashed), and 
$\kappa=\infty$ (dotted) are shown.} \label{fig:combined}
\end{center}
\end{figure}

To include both types of scattering effects, we neglect
interference processes between electron-electron collisions and
impurity scattering entirely, and approximate the total scattering
rate at the node by
\begin{equation}
\Gamma_{\rm tot}=\Gamma_{\rm el}+\Gamma_{\rm inel},
\label{twentyseven}
\end{equation}
i.e. higher order processes like the influence of the inelastic
scattering on the elastic scattering and vice versa are neglected.
However, this is partially taken
into account since we have chosen the parameters for the elastic and
inelastic scattering on a phenomenological basis.

In Fig.~\ref{fig:combined}, we plot the on-shell total
scattering rate $\Gamma_{\rm tot}$ of a nodal quasiparticle as a
function of both temperature and energy.  It is seen that the
generic features, for reasonable assumptions about the magnitudes
of the impurity scattering rates and ranges, are as follows.  As a
function of temperature, one expects the nodal scattering rate to
collapse at $T_c$, and to obey a $T^3$ dependence at the lowest
temperatures, whose coefficient is determined by both the elastic and
inelastic processes.
On the other hand, the low {\it energy} dependence
of the scattering rate at low temperatures is dominated
by the quasi-linear $\omega$ dependence of the elastic scattering.
The generic result appears to be a
quasilinear low-energy behavior for weak out of plane impurities,
where the slope of this result is unrelated to the slope of the
high-energy quasilinear behavior determined by spin-fluctuation
scattering.  A very low-energy $(\omega \sim 1-2$meV) flattening or upturn due to
unitary scatters ($\sum_\k G\simeq \omega$) at low energies in
Eq.~\eqref{three}, is expected to be present in principle. This contribution 
is likely to be
negligible in BSCCO due to the small concentration of native
planar defects in current samples, and the fact that the expected
impurity bandwidth is comparable to the energy resolution in ARPES
experiments.

\section{Conclusions}

Understanding the lifetime of nodal quasiparticles is a
fundamental problem of cuprate physics which must be solved if we
are to agree that a $d$-wave BCS theory describes the optimally to
over-doped superconducting state, the only part of the phase
diagram apparently susceptible to semiquantitative description at
this time.  In addition, the fact that nodal quasiparticles appear
to be robust even for strongly underdoped samples may be an
important clue to the physics of the pseudogap, and a theory of
the nodal states within BCS which works at optimal doping may help
us to understand this clue. Finally, the lifetime of the nodal
quasiparticles determines thermal and microwave conductivity as well
as photoemission lineshapes and one would like to have a unified
description of these quantities within a single model.
Here, we have calculated the on-shell $\omega=\varepsilon_k$
nodal quasiparticle scattering rate which enters in determining the MDC linewidth.
Using a simple model which parameterizes the forward elastic scattering in terms of a
range $\kappa^{-1}$, a normal state scattering rate $\Gamma_0$, and an
inelastic spin-fluctuation
scattering parameterized by an RPA form with an effective Coulomb coupling $U$, we have
studied the temperature and $\omega$-dependence of the scattering rate.

For a quasiparticle at the nodal momentum $k_N$, as $T$ decreases
below $T_c$, the elastic scattering rate decreases from its normal
state value $\Gamma_0$. As the gap opens, it reaches a smaller
value $\Gamma_{\rm el} (T)$ determined by the forward scattering
parameter $\kappa$ and the renormalized gap velocity $\tilde v_2$.
This temperature dependence of the elastic scattering rate arises
from the temperature dependence of the gap amplitude which
controls the available phase space for scattering, suppressing it
as the gap opens. Because the scattering rate depends
exponentially on the gap, which  opens rapidly in BSCCO, the
elastic scattering rate has a cusp-like  onset at $T_c$,
decreasing rapidly to its low temperature value. The inelastic
contribution to scattering rate also decreases as the gap opens,
with a cusp, and then decreases as $(T/\Delta_0)^3$ at low
temperatures.

At low reduced temperatures, the $\omega$-dependence of the elastic scattering rate can exhibit a
quasi-linear behavior, varying as
$
\frac{\omega}{6} \left(\frac{\tilde v_2\kappa}{\Gamma_0}\right)
$
if the scattering is forward or as $\Gamma_0\frac{\omega}{\Delta_0}$ if the scattering is
more isotropic.
The inelastic scattering
rate initially increases as $\omega^3/\Delta^2_0$ so that there is an energy
beyond which the inelastic scattering becomes dominant.  For
energies greater than of order $3\Delta_0$, the inelastic scattering rate crosses over to a
quasi-linear $\omega$-dependence with a slope of order one.

Although the calculations we have presented here are
straightforward, at the present time the experimental situation
regarding the direct measurement of the nodal scattering rate by
ARPES is somewhat uncertain. There are some claims in the
literature that the rate collapses in the SC state \cite{Kam00},
as found theoretically here, and some that marginal Fermi liquid
linear behavior consistent with a quantum critical point persists
down to the lowest temperatures \cite{Val00}. As samples and
resolution of the ARPES technique improve, we expect this
discrepancy to be resolved and our prediction for the nodal
quasiparticle MDC width to be testable.

%

\acknowledgments

The authors are grateful for discussions
with D.~Dessau, A.~Fujimori, A.~Ino, P.~Johnson, and Z.-X.~Shen. We would 
like to thank A. Chubukov for his comments regarding the second order 
on-shell inelastic scattering rate. Work was
partially supported by ONR grant N00014-04-0060 and NSF
DMR02-11166.

\end{document}